\documentclass[aps,prl,twocolumn, superscriptaddress]{revtex4}

\usepackage{graphicx}
\usepackage{dcolumn}
\usepackage{bm}
\usepackage{wrapfig}
\usepackage{amssymb}
\usepackage{amsmath}
\usepackage{color}
\usepackage{multirow}
\usepackage{pdfsync}
\usepackage{color}
\usepackage[normalem]{ulem}

\def\be{\begin{equation}}
\def\ee{\end{equation}}
\def\bea{\begin{eqnarray}}
\def\eea{\end{eqnarray}}
\def\la{\langle}
\def\ra{\rangle}
\definecolor{darkgreen}{rgb}{0, 0.4, 0}

\newcommand{\ie}{{\it{i.e.~}}}

\begin{document}

\title{Maximal quantum randomness in Bell tests}

\author{Chirag Dhara}
\email{chirag.dhara@icfo.es} \affiliation{ICFO-Institut de
Ciencies Fotoniques, Mediterranean Technology Park, 08860
Castelldefels (Barcelona), Spain}
\author{Giuseppe Prettico}
\email{giuseppe.prettico@icfo.es} \affiliation{ICFO-Institut de
Ciencies Fotoniques, Mediterranean Technology Park, 08860
Castelldefels (Barcelona), Spain}
\author{A. Ac\'in}
\affiliation{ICFO-Institut de Ciencies Fotoniques, Mediterranean
Technology Park, 08860 Castelldefels (Barcelona), Spain}
\affiliation{ICREA-Instituci\'o Catalana de Recerca i Estudis
Avan\c cats, Lluis Companys 23, 08010 Barcelona, Spain}

\date{\today}

\begin{abstract}
The non-local correlations exhibited when measuring entangled particles can be used to certify the presence of genuine randomness in Bell experiments. While non-locality is necessary for randomness certification, it is unclear when and why non-locality certifies maximal randomness. We provide a simple argument to certify the presence of maximal local and global randomness based on symmetries of a Bell inequality and the existence of a unique quantum probability distribution that maximally violates it. We prove the existence of N-party Bell tests attaining maximal global randomness, that is, where a combination of measurements by each party provides N perfect random bits.
\end{abstract}


\maketitle

{\it Introduction.}
Quantum theory radically departs from
classical theory in many aspects. Quantum theory, for instance,
predicts correlations among distant
non-communicating observers that cannot be reproduced classically.
These correlations are termed non-local and violate those
conditions known as Bell inequalities that, in contrast, are satisfied by
classically correlated systems~\cite{bell64}. Quantum theory also
incorporates a form of randomness in its framework that does not
have a classical counterpart. There is no true randomness in
Newtonian physics, as the complete knowledge of initial conditions
along with interactions of a system allows one to predict its
future dynamics deterministically. As well known however,
predictions in quantum systems are necessarily probabilistic. Since violation of Bell inequalities implies that quantum theory cannot be explained by local deterministic theories, the probabilistic nature must arise from intrinsic randomness. Hence, the violation of a
Bell inequality certifies the existence of genuine randomness (for
recent developments, see~\cite{us} and references therein).

The relation between non-locality and randomness has attracted the
interest of physicists since the very inception of quantum theory.
While earlier motivated by its foundational implications, it has
acquired a practical aspect due to the rapid developments in
quantum information and computation. Randomness constitutes a
valuable information resource, with applications ranging from
cryptographic protocols and gambling to numerical simulations of
physical and biological systems. Recently, tools to certify and
quantify the presence of randomness in Bell tests have been
presented in~\cite{pironio}. An important advantage of this
approach is that it is derived in the device-independent scenario,
where it is possible to characterize the system from
an input-output perspective without regard for its internal working. While, as said, we now have tools to
link quantum randomness and non-locality, we are still far from
understanding the exact relation between these two quantum
properties. For instance, there are situations in which a
probability distribution with maximal non-locality does not
necessarily contain maximal randomness. Even more counter
intuitively, distributions with arbitrarily small non-locality can
contain almost maximal randomness in some cases~\cite{Toni}. Along
this direction, identifying those quantum set-ups, namely Bell
tests, which offer the highest possible randomness would be a
highly desirable result, both from a fundamental and practical
point of view. This is the main goal of the present work.

It is worth illustrating our motivations with an example. Consider
the standard Clauser-Horne-Shimony-Holt (CHSH)
inequality~\cite{chsh}, $I_{CHSH}=\la A_1B_1\ra +\la A_1B_2\ra
+\la A_2B_1\ra -\la A_2B_2\ra$. At the point of maximal quantum
violation, any measurement output by any of the parties provides a
perfect random bit. That is, the corresponding probability
distribution contains {\it locally} the maximum possible of one
bit of randomness for every party and every measurement setting.
However, there are strictly less than $2$ random bits {\it
globally}, as any pair of local measurements gives correlated
results. Now, consider the following modification of the CHSH
inequality, $I_{\eta}=\la A_1B_1\ra +\la A_1B_2\ra +\la A_2B_1\ra
-\la A_2B_2\ra+\eta \la A_1\ra$. At the point of maximal quantum
violation, only the measurement $A_2$ defines a perfect random
bit~\cite{Toni}. Why this setting and not the others? Why all of
them in the case of CHSH? More in general, when can we expect
maximal local, and global randomness in a Bell test?

Our main result is a simple method to infer when and which
settings in a Bell test certify perfect random bits. Given a Bell
inequality, our method (i) assumes that the quantum probability
distribution attaining its maximal violation is unique and (ii)
exploits the symmetries of the inequality. We show how this method
reproduces all known results relating Bell tests and maximal
randomness. Moreover, based on our construction, we provide Bell
tests certifying the maximal global randomness in a robust manner,
that is, Bell tests for which there exist measurements by the $N$
parties providing $N$ random bits.  We also provide a geometric
interpretation of our findings. Finally, we discuss the existence of uniqueness and show that it is known to exist in several important cases either analytically or from numerical computation.

We start by explaining our notation and stating the basic
definitions we use in the text.

{\it Bell tests and quantum distributions.} We denote by $(N,M,d)$
a standard Bell experiment consisting of $N$ separated and
non-communicating parties, where each of them can perform $M$
local measurements of $d$ outcomes. By repeating the experiment,
it is possible to assign a probability distribution
$P(a_1,\dots,a_N|x_1,\dots,x_N)$, where $a_i$ is the outcome of a
measurement $x_i$ by party $1 \leq i\leq N$. We often consider
cases with dichotomic measurements \ie $d=2$. In this case, we can
use the following useful parametrization,
\begin{widetext}
\bea \label{ProbDistrCorr} P({\bf a}|{\bf
x})=\frac{1}{2^N}\left(1+\sum_{i=1}^N a_i\la A_i\ra+\sum_{i<j} a_i
a_j\la A_i A_j\ra+\sum_{i<j<k} a_i a_j a_k\la A_i A_j
A_k\ra+\dots+a_1a_2\dots a_N\la A_1 A_2\dots A_N\ra\right) . \eea
\end{widetext}
Here, measurement outputs are labeled by $\pm 1$ and $\la
A_i\ldots A_j\ra$ are the standard correlators $\la A_i\ldots
A_j\ra=Pr(A_i \ldots A_j=+1)-Pr(A_i\ldots A_j=-1)$.
\vspace{0.1cm}

{\it Randomness.} We follow~\cite{pironio,Toni} and adopt an operational approach where
randomness is related to the probability of correctly guessing the
outcome of some joint measurement, ${\bf x}=(x_1,x_2,\dots,x_N)$.
We use the {\it guessing probability},
$P_G(P;\textbf{x})=\max_{{\bf a}}P({\bf a}|{\bf x})$, where ${\bf
a} =(a_1,a_2,\dots,a_N)$. The proper measure of intrinsic randomness requires optimizing over all realizations
of the observed correlations $G(P;{\bf x})=\max \sum_i
\lambda_iP_G(P_i;\textbf{x})$, where the maximization is over all
convex decompositions $P({\bf a} |{\bf x})=\sum_i
\lambda_iP_i({\bf a} |{\bf x})$. It is convenient to express the
randomness in bits with the {\it min-entropy}, $H_{\infty} (P;{\bf
x})=-\log_2G(P;{\bf x})$. Note that in a general $(N,M,d)$
scenario there can be at most $\log_2d$ bits of local and $N\log_2
d$ bits of global randomness at any given round of the experiment.
For a given ${\bf x}={\bf x_0}$, maximal randomness is obtained
from a uniform distribution $P({\bf a}|{\bf
x_0})=1/d^N,\forall\,\textbf a$. When $d=2$, this occurs if, and
only if, all the correlators appearing in~(\ref{ProbDistrCorr})
are zero.
\vspace{0.1cm}

{\it Maximal randomness certification.} The main
result of our work is a simple method to infer when some settings
in a Bell test can provide maximal randomness. We assume in what
follows that the quantum distribution attaining the maximal
quantum violation of the Bell inequality is unique (discussed later). Under this assumption, we
show how symmetries in the Bell inequality under permutation of
measurement results, possibly together with permutations of
measurement settings, lead to maximal randomness. Our method,
then, can be summarized as follows: {\it uniqueness plus
symmetries implies maximal randomness}.

To illustrate our method, it is worth reexamining the examples
given above. Consider again the CHSH inequality and denote by
$\mathcal{P}^*$ the distribution attaining its maximal quantum
violation, namely $I_{CHSH}(\mathcal{P}^*)=2\sqrt{2}$. Note that
in this case, this distribution is known to be unique~\cite{tsi1}.
The symmetry transformation $\mathcal{T}_s$: $a_{1,2}\mapsto
-a_{1,2}$ and $b_{1,2}\mapsto -b_{1,2}$ flips the signs of all the
one-body correlators, $\la A_i\ra$ and $\la B_j\ra$, while keeps
unchanged all two-body correlators, $\la A_iB_j\ra$. Applying
$\mathcal{T}_s$ to $P^*$ we obtain a new distribution
$\mathcal{T}_s(\mathcal{P}^{*})=\mathcal{P}^{**}$ with
\begin{align}\label{marginals}
\la A_i\ra^{**}=-\la A_i\ra^{*},\qquad  & \la B_j\ra^{**}=-\la
B_j\ra^{*} ,
\end{align}
and that also maximally violates CHSH. Because of the uniqueness
of the distribution, $\mathcal{P}^{*}=\mathcal{P}^{**}$ and all
one-body correlators~(\ref{marginals}) must be zero, which
certifies $1$ bit of \emph{local} randomness (for both parties).
Moving to $I_{\eta}$, the transformation $a_2\mapsto -a_2$, $B_1
\leftrightarrow B_2$, flips the value of $\la A_2\ra$ without
changing the value of $I_\eta$. Under the assumption of
uniqueness, this proves that the setting $A_2$ is fully random. A
little thought shows that it is impossible to construct similar
transformations for the other local measurements. Our argument,
then, easily reproduces the known results for these two
inequalities.

As mentioned, our method applies to any Bell inequality with
symmetries. The previous argument for the CHSH inequality can be
easily generalized to all the chained inequalities of
Refs.~\cite{Braun,bkp}. Under the assumption of uniqueness, these
inequalities always certify $1$-dit of local randomness. The
chained Bell inequalities  can be compactly represented
as~\cite{bkp}:
\be \label{chain} C^M_d = \sum_{i=1}^M \la [A_i-B_i]_d\ra + \la
[B_i-A_{i+1}]_d\ra \geqslant d-1 \ee where $A_i$, $B_j\in
\{0,\dots,d-1\}$ are measurement choices for Alice and Bob and
$A_{M+1}=A_1+1$. The square brackets denote sum modulo $d$.

Let $P$ attain the quantum maximum of $C_d^M$. The transformation
$\cal{T}$: $a_i\mapsto a_i+1$ and $b_i\mapsto b_i+1$ for every $i$
changes the value of the marginal distributions of Alice and Bob
but leaves the terms in $C_d^M$ unchanged. Applying $\cal{T}$ to
$P$ and assuming it to be unique, it follows that all local
distributions of Alice and Bob must be uniform. In other words,
the chained inequality certifies $\log_2 d$ bits of local
randomness for every measurement by each party.
\vspace{0.1cm}

{\it Bell tests attaining maximal global randomness.} A natural
open question is whether there exist Bell tests in the $(N,M,d)$
scenario that allow certifying the maximal possible randomness,
namely $N\log_2 d$ bits.  Some progress on this question was
obtained in~\cite{Toni}, where it was shown how to get arbitrarily
close to two random bits in the $(2,2,2)$ scenario. However the corresponding correlations are non-robust. Here, we show how our method can be easily applied to
design Bell tests allowing maximal randomness certification in a
robust manner.

We start with the bipartite case. Maximal global randomness is
impossible in the CHSH case, as at the point of maximal violation
all settings are correlated. Maximal global randomness, however,
can be certified as soon as another measurement is included. More
in general, consider the chained inequalities for an odd number of
two-outcome measurements. We move to the notation $a_i,b_j = \pm
1$ and reexpress (\ref{chain}) as follows: \bea \label{chainM2}
C^M_2=\left | \sum_{i=1}^M \la A_i B_i\ra + \sum_{i=1}^{M-1} \la
A_{i+1} B_i\ra - \la A_{1} B_M\ra \right| \eea where $A_i$, $B_j
=\pm 1$. Let $M=2k +1$. As above, we consider a transformation
leaving $C^M_2$ unchanged but under which $\la A_{1} B_{k+1}\ra \mapsto -\la
A_{1} B_{k+1}\ra$. Such a transformation is: $\cal{T}$: $a_1\mapsto -a_1$, $B_{1+i}
\leftrightarrow B_{M-i}$, $A_{2+i} \leftrightarrow A_{M-i}$
$\forall i \ 0\leq i\leq k-1$.
Assuming that the distribution maximally violating (\ref{chainM2})
is unique leads to  $\la A_{1} B_{k+1}\ra=0$. The previous results
show that $\la A_1\ra=0=\la B_{k+1}\ra$. These together certify 2
bits of global randomness for $(A_1,B_{k+1})$. Similar arguments
certify maximal randomness in all inputs of the form
$(A_{l},B_{k+l})$ $\forall$ $1\leq l\leq k$.  Analogous to the case for CHSH, maximal
randomness cannot be certified  for those measurement combinations
appearing in the chained inequality, as they display non-zero
correlations. The previous results rely on the assumption of
uniqueness, which is unknown for the case of the chained
inequality with $M>2$. We then follow~\cite{pironio} and apply the
techniques in~\cite{npa} to get an upper bound on the randomness
of $(A_1,B_2)$ for the chained inequality with 3 measurement
settings. The obtained results corroborate the presence of maximal
global randomness, up to numerical accuracy.

We now move to the multipartite case. More precisely, we consider
the Mermin inequalities~\cite{mermin} and prove that they allow
certifying up to $N$ bits of global randomness for arbitrary odd
$N$. Mermin inequalities of $N$ parties
are defined recursively as, \be \label{mermin_recur}
M_N=\frac{1}{2}M_{N-1}(A_N+A_N')+\frac{1}{2}M_{N-1}'(A_N-A_N') \ee
where $M_2$ is the CHSH inequality and $M_{N-1}'$ is obtained from
$M_{N-1}$ by exchanging all $A_j$ and $A_j'$.

Let $M_N$ denote a Mermin inequality of $N=2J+1$ sites. Party $i$,
with $i=1,\ldots,N$ has a choice between two dichotomic
measurements, $A_i$ and $A_i'$. It is easily checked that for odd
$N$, $M_N$ contains only full correlators with an odd number of
primes. We show, using symmetry arguments, that at the point of maximal
quantum violation every correlator  $\la A_i\ldots A_j\ra$
(involving an arbitrary number of measurements) that does not
appear in $M_N$ is identically zero. This automatically implies
that any combination of $N$ settings not appearing in the
inequality define $N$ random bits.

To see this, first take a specific $N$-body correlator not
appearing in $M_N$, $\la X_1X_2 \dots X_{N}\ra$ where $X_i =A_i$
or $A_i'$ but such that the total number of primed $A$ is an even
number. Denote the outcome of $X_i$ by $x_i$. Choose any of the
parties, say the first one, and denote by $Corr(X_1)$ the set of
all correlators of arbitrary size containing $X_1$ plus possibly
other settings $X_i$ with $i>1$. We would like to show that every
element belonging to $Corr(X_1)$ is equal to zero for the unique
distribution maximally violating the inequality. Let us consider
the transformation $\mathcal{S}_1:\{x_{1}\mapsto -x_{1}$, and
$x_{j}$ untouched $\forall j >1\}$. This maps $Corr(X_{1})\mapsto
-Corr(X_{1})$. The terms in $M_N$ remains unchanged if we
complement $\mathcal{S}_1$ with $\mathcal S'_1:\{x_{j}' \mapsto
-x_{j}' \forall j>1 \}$, where we use $(A'_i)'=A_i$.
In fact, note that for the original even primed term we started
with, $\mathcal S'_1\circ\mathcal{S}_1\la X_1X_2 \dots
X_{N}\ra=-\la X_1X_2 \dots X_{N}\ra$. The Mermin inequality
consists only of odd-parity full-correlators. Any such a term can
be obtained from $\la X_1X_2 \dots X_{N}\ra$ by swapping inputs at
an odd number of places. However, the transformation $\mathcal
S'_1\circ\mathcal{S}_1$ is such that at every site, either the
outcome of $A_i$ or $A_i'$ flips sign but not both. Hence,
$\mathcal S'_1\circ\mathcal{S}_1$ applied on any correlator
obtained by an odd number of local swaps on $\la X_1X_2 \dots
X_{N}\ra$ gains an additional factor of $-1$ {\it for each swapped
site} relative to $\mathcal S'_1\circ\mathcal{S}_1\la X_1X_2 \dots
X_{N}\ra$. Thus, $M_N$ remains unchanged. It remains to study the
effect of $\mathcal S'_1$ on $Corr(X_{1})$. Since $X_j' \notin
Corr(X_{1})$, this set is unmodified under $\mathcal S'_1$, so
$\mathcal S'_1\circ\mathcal{S}_1$ maps $Corr(X_{1})\mapsto
-Corr(X_{1})$. We then conclude from uniqueness that all the correlators in
$Corr(X_{1})$ must be zero. The same argument can be run for any
party, and then for any full-correlator with an even number of
primes, proving the result.

Before concluding this part, it is worth mentioning that similar
arguments when applied to the Mermin inequality for even $N$ allow
certifying $(N-1)$ bits of randomness.
\vspace{0.1cm}

{\it Geometric interpretation.} The previous argument crucially
relies on the assumption that there is a unique quantum
distribution attaining the maximal violation of a given Bell
inequality. For some cases, such as Mermin  $(N,2,2)$, this uniqueness has
been proven~\cite{torsten,werner} and, then, it is no longer an
assumption. For the chained inequality, we have numerical evidence
using the techniques from~\cite{npa} that the distribution
saturating it is unique in the $(2,3,2)$ and  $(2,4,2)$ cases.

From a geometrical point of view, it is natural to expect that the
maximal violation of a generic Bell inequality is attained by a
unique point. The set of quantum correlations defines a convex set
in the space of probability distributions
$P(a_1,\dots,a_N|x_1,\dots,x_N)$. A Bell inequality is a
hyperplane in this space. The maximal quantum violation
corresponds to the point in which the hyperplane, \ie the Bell
inequality, becomes tangent to the set of quantum correlations.
Since the set is convex, this point is expected to be unique, in
general. Of course, there may be situations for which this is not
true. So far the only exceptions we have found from numerics are
for {\it lifted} Bell inequalities. A tight Bell inequality of a
smaller space can be lifted in a sense made precise in
\cite{lifted_pironio} to a tight Bell inequality in a higher
space, either with more parties, measurements or outcomes. For
example, $(CHSH-2)_{AB}\otimes C_1\leq 0$ is a tight Bell
inequality of $(3,2,2)$ in which party $C$ only applies one
measurement. It is easy to see that there are several quantum
realizations attaining the maximal violation of this inequality.
However, it may be argued that these Bell inequalities should be
properly be considered as belonging to a lower dimensional space.
\begin{figure}[htbp]
\begin{center}
\includegraphics[width= 8.5cm]{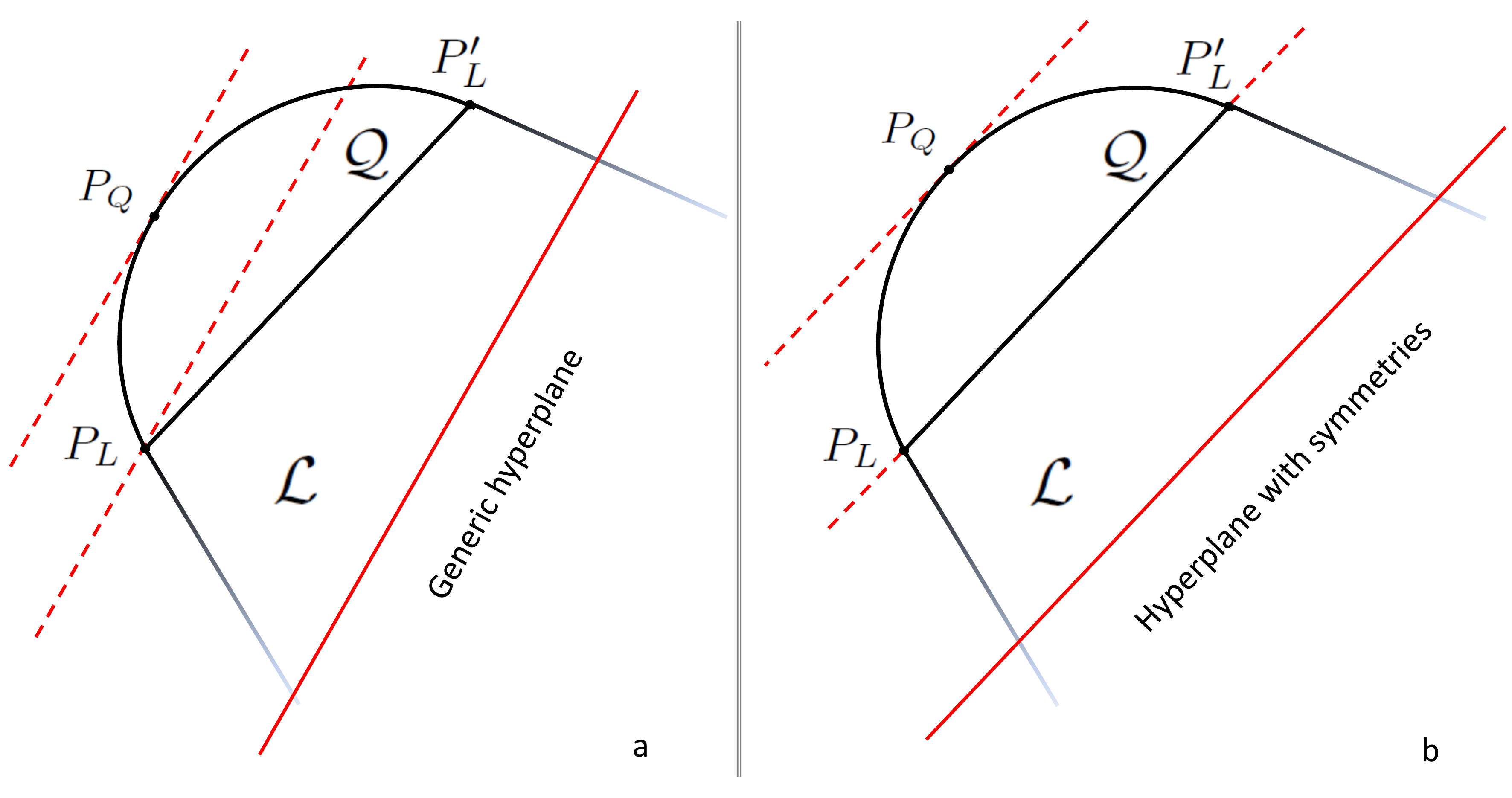}
\caption{a) A generic hyperplane generally does not have symmetries and has a unique maximum in both the local and the quantum sets. b) A hyperplane with symmetries (such as the CHSH) precludes uniqueness in the local set but still allows for a unique maximum in the quantum set.} \label{generic}
\end{center}
\end{figure}

One should, however, be careful when following this geometrical
intuition. Note that the previous argument does not make use of
any quantum property. In fact, the set of classical correlations
is also convex and, thus, a generic hyperplane is expected to
become tangent at a unique extremal point, see
Fig.~\ref{generic}a. However, randomness cannot be
certified by classical correlations. The reason is that our method
applies only to Bell inequalities that are symmetric under
permutation of some of the measurement results, possibly assisted
by permutations of measurements. It is easy to see that, within
the local set, any symmetry under permutations of the results can
be immediately used to construct another extremal and
deterministic point saturating the inequality as in Fig.~\ref{generic}b.

How do these considerations extend to general non-signalling
correlations? While this is beyond the scope of the present work,
we just pointed out here that the chained inequality allows
certifying at most one bit of global randomness~\cite{lluis}, as opposed to the
two bits in the quantum case. This implies that there is more than
one non-signalling point saturating the inequality. Understanding
why randomness certification, based on uniqueness and symmetries,
behaves so differently in the quantum set is an interesting
question that deserves further investigation. From a speculative
point of view, the fact that the quantum set is not a polytope, as
opposed to the set of classical and non-signalling correlations,
may play a key role in these considerations.

{\it Conclusions.} Our
argument is based on the simultaneous existence of uniqueness and symmetries. While
in the classical case the needed symmetries immediately break the
uniqueness of the maximal violation, this is no longer the case
for quantum correlations, as implied by our results. Furthermore, we are
yet to find an example where results from our symmetry arguments are in
contradiction with numerical results where such computation was possible.
For instance, for the I3322~\cite{Froissart, Sliwa, Collins} inequality, there are no symmetry
arguments possible in order to certify maximal local randomness and, in
fact, the known maximal quantum violation of the inequality gives
non-uniform marginals\cite{vertesi}.

While our simple recipe does not constitute a formal proof of randomness unless
uniqueness is proven it still turns
out to be very useful to find the right Bell inequalities and measurements
allowing maximal randomness certification. Indeed, the results
derived following our method can later be confirmed using the
techniques from~\cite{pironio,npa}. In this sense, we are not
aware of any Bell test leading to maximal randomness, local or
global, that cannot be explained using our method. Our findings
indicate that settings not appearing in the Bell inequality may
have more global randomness than those appearing in the
inequality. Moreover, using our method, we easily demonstrated the
existence of Bell tests allowing maximal global randomness.
Finally, our work opens new perspectives on the relation between
randomness and non-locality that deserve further investigation.

\vspace{0.1cm}

\begin{acknowledgements}
We acknowledge support from the ERC Starting Grant PERCENT, the EU Projects Q-Essence and QCS, the Spanish MICIIN through a Juan de la Cierva grant and the Spanish FPI grant,  an FI Grant of the Generalitat de Catalunya and projects FIS2010-14830, Explora-Intrinqra, CHIST-ERA DIQIP.
\end{acknowledgements}


\end{document}